\documentclass[twocol]{ametsoc}
\usepackage{color,soul}
\usepackage{amsmath,amssymb}
\usepackage{cancel}

\journal{jas}
\bibpunct{(}{)}{;}{a}{}{,}

\title{An Eddy-Zonal Flow Feedback Model for Propagating Annular Modes}

    \authors{Sandro W. Lubis\correspondingauthor{6100 Main Street, MS-321, Houston, Texas, USA} and Pedram Hassanzadeh}
    \affiliation{Rice University, Houston, Texas, USA}
    \email{slubis@rice.edu, pedram@rice.edu}

\abstract{The variability of the zonal-mean large-scale extratropical circulation is often studied using individual modes obtained from empirical orthogonal function (EOF) analyses. The prevailing reduced-order model of the leading EOF (EOF1) of zonal-mean zonal wind, called the annular mode, consists of an eddy-mean flow interaction mechanism that results in a positive feedback of EOF1 onto itself. However, a few studies have pointed out that under some circumstances in observations and GCMs, strong couplings exist between EOF1 and EOF2 at some lag times, resulting in decaying-oscillatory, or propagating, annular modes. Here, we introduce a reduced-order model for coupled EOF1 and EOF2 that accounts for potential cross-EOF eddy-zonal flow feedbacks. Using the analytical solution of this model, we derive conditions for the existence of the propagating regime based on the feedback strengths. Using this model, and idealized GCMs and stochastic prototypes, we show that cross-EOF feedbacks play an important role in controlling the persistence of the annular modes by setting the frequency of the oscillation. We find that stronger cross-EOF feedbacks lead to less persistent annular modes. Applying the coupled-EOF model to the Southern Hemisphere reanalysis data shows the existence of strong cross-EOF feedbacks. The results highlight the importance of considering the coupling of EOFs and cross-EOF feedbacks to fully understand the natural and forced variability of the zonal-mean large-scale circulation.}

\begin{document}
\maketitle


\section{Introduction}
\label{sec:1}

At the intraseasonal to interannual time scales, the variability of the large-scale atmospheric circulation in the mid-latitudes of both hemispheres is dominated by the ``annular modes'', which are usually defined based on empirical orthogonal function (EOF) analysis of zonal-mean meteorological fields (e.g., \citealt{Kidson1988,ThompsonWallace1998,ThompsonWallace2000,Lorenz2001,Lorenz2003,ThompsonWoodworth2014,ThompsonLi2015}). The barotropic annular modes are often derived as the first (i.e., leading) EOF (EOF1) of zonal-mean zonal wind, which exhibits a dipolar meridional structure and describes a north-south meandering of the eddy-driven jet. Note that in this paper, the focus is on the barotropic annular modes, hereafter simply called annular modes (see \citealt{ThompsonWoodworth2014,ThompsonBarnes2014}, and \citealt{ThompsonLi2015} for discussions about the ``baroclinic annular modes''). The second EOF of zonal-mean zonal wind (EOF2) has a tripolar meridional structure centered on the jet, describing a strengthening and weakening  of the eddy-driven jet (i.e., jet pulsation). By construction, EOF1 and EOF2 (and any two EOFs) are orthogonal and their associated time series (i.e., principal components, PCs), sometimes called zonal index, are independent at zero time lag.

The persistence of the annular mode (EOF1) and its underlying dynamics have been the subject of extensive research and debate in the past three decades \citep[e.g.,][]{Robinson1991,Branstator1995,FeldsteinLee1998,Robinson2000,Lorenz2001,Lorenz2003,Gerber2007,Gerber2008,ChenPlumb2009,SimpsonShepherd2013,Zurita2014,Nie2014,Byrne2016,MaHassanzadehKuang2017,HassanzadehKuang2019}. Many of the aforementioned studies have pointed to a positive eddy-zonal flow feedback mechanism as the source of the persistence: The zonal wind and temperature anomalies associated with the annular mode (EOF1) modify the generation and/or propagation of the synoptic eddies at the quasi-steady limit (greater than 7 days) in such a way that the resulting eddy fluxes reinforce the annular mode (see \citet{HassanzadehKuang2019} and the discussion and references therein). Most notably, \cite{Lorenz2001} developed a linear eddy-zonal flow feedback model (LH01 model hereafter) for the annular modes by regressing the anomalous eddy momentum flux divergence onto the zonal index of EOF1 ($z_1$) and interpreting correlations between $z_1(t)$ and regressed momentum flux divergence ($m_1(t)$) at long lags (greater than 7 days) as evidence for eddy-zonal flow feedbacks, i.e., feedbacks of EOF1 onto itself. \citet{Lorenz2001} developed a similar model, separately, for EOF2 and found, respectively, positive and weak eddy-zonal flow feedbacks for EOF1 and EOF2, respectively, consistent with the longer persistence of EOF1 compared to EOF2. Such single-EOF eddy-zonal flow feedback models have been used in most of the subsequent studies of the annular modes \citep[e.g.,][]{Lorenz2003, SimpsonShepherd2013,Lorenz2014,Robert2017,MaHassanzadehKuang2017,BoljkaShepherd2018,HassanzadehKuang2019,Lindgren2020}.

While EOF1 and EOF2 are independent at zero lag, a few previous studies have pointed out that these two EOFs can be correlated at long lags (e.g., greater than 10 days), and that in fact the combination of these two leading EOFs represents coherent meridional propagations of the zonal-mean flow anomalies. Such propagating regimes have been observed in both hemispheres in reanalysis data \citep[e.g.,][]{Feldstein1998,FeldsteinLee1998,SheshadriPlumb2017}. Anomalous poleward propagation of zonal wind typically emerges in low latitudes and mainly migrate poleward over a few months, although non-propagating regimes can also appear in some instances (see Fig.~1 of \citealt{SheshadriPlumb2017} and Fig.~\ref{fig:6} in this paper). Similar behaviors have also been reported by in general circulation models (GCMs) (e.g., \citealt{JamesDodd1996,SonLee2006,SonLee2008,SheshadriPlumb2017}). \cite{SonLee2006} found that the leading mode of variability in an idealized dry GCM can be either the propagating or non-propagating regime depending on the details of thermal forcing imposed in the model. They also found that unlike the non-propagating regimes, the $z_1$ and $z_2$ of the propagating regimes are strongly correlated at long lags, peaking around $50$~days (see their Fig.~3; also Figs.~\ref{fig:4}b of the present paper). Furthermore, \cite{SonLee2006} reported that non-propagating regimes are often characterized by a single time-mean jet with a dominant EOF1 (in terms of the explained variance) while the propagating regimes are characterized by a double time-mean jet in the mid-latitudes with the variance associated with EOF2 being at least half of the variance of EOF1. Furthermore, \cite{SonLee2008} found that the $e$-folding decorrelation time scale of $z_1$ in the propagating regime to be much shorter than that of the non-propagating regime. The long $e$-folding decorrelation time scales for the annular modes in the non-propagating regime were attributed to an unrealistically strong positive EOF1-onto-EOF1 feedback, while the reason behind the reduction in the persistence of the annular modes in the propagating regime remained unclear.

More recently, \citet{SheshadriPlumb2017} presented further evidence for the existence of propagating and non-propagating regimes and strong lagged correlations between $z_1$ and $z_2$ in reanalysis data of the Southern Hemisphere (SH) and in idealized GCMs. Moreover, they elegantly showed, using a principal oscillation patterns (POP) analysis \citep{Hasselmann1988,Penland1989}, that EOF1 and EOF2 are in fact manifestations of a single, decaying-oscillatory coupled mode of the dynamical system. Specifically, they found that EOF1 and EOF2 are, respectively, the real and imaginary parts of a single POP mode, which describes the dominant aspects of the spatio-temporal evolution of zonal wind anomalies. \citet{SheshadriPlumb2017} also showed that in the propagating regime, the auto-correlation functions of $z_1$  and $z_2$ decay non-exponentially.

Given the above discussion, a single-EOF model is not enough to describe a propagating regime because the EOF1 and EOF2 in this regime are strongly correlated at long lags and that the auto-correlation functions of the associated PCs do not decay exponentially (but rather show some oscillatory behaviors too). From the perspective of eddy-zonal flow feedbacks, one may wonder whether there are cross-EOF feedbacks in addition to the previously studied EOF1 (EOF2) eddy-zonal flow feedback onto EOF1 (EOF2) in the propagating regime. In cross-EOF feedbacks, EOF1 (EOF2) changes the eddy forcing of EOF2 (EOF1) in the quasi-steady limit. Therefore, there is a need to extend the single-EOF model of LH01 and build a model that includes, at a minimum, both leading EOFs and accounts for their cross feedbacks. The objective of the current study is to develop such a model and to use it to estimate effects of the cross-EOF feedbacks on the variability of propagating annular modes.

The paper is structured as follows: Section~\ref{sec:2}  compares the characteristics of $z_1$, $z_2$, $m_1$, and $m_2$ for the non-propagating and propagating annular modes in reanalysis and idealized GCMs. In Section~\ref{sec:3} , we develop a linear eddy-zonal flow feedback model that accounts for cross-EOF feedbacks, validate the model using synthetic data from a stochastic prototype, discuss the key properties of the analytical solution of this model, and apply this model to data from reanalysis and an idealized GCM. The paper ends concluding remarks in Section~\ref{sec:4}.

\section{Propagating annular modes in an idealized GCM and reanalysis}
\label{sec:2}

In this section, we will examine the basic characteristics and statistics of propagating annular modes in an idealized GCM (the dry dynamical core) and reanalysis. We focus on the southern annular mode, which makes it easier to compare the results of the reanalysis and the idealized aquaplanet GCM simulations. We will start with the idealized GCM to demonstrate the characteristics of the propagating and non-propagating annular modes. 

\subsection{An idealized GCM: The dry dynamical core}
\label{sec:21}

We use the Geophysical Fluid Dynamics Laboratory (GFDL) dry dynamical core GCM. The GCM is run with a flat, uniform lower boundary (i.e., aquaplanet) with T63 spectral resolution and 40 evenly spaced sigma levels in the vertical for 50000-day integrations after spinup. The physics of the model is based on \cite{Held_Suarez1994}, an idealized configuration for generating a realistic global circulation with minimal parameterization \citep{Held2005,Jeevanjee2017}. All diabatic processes are represented by Newtonian relaxation of the temperature field toward a prescribed equilibrium profile, and Rayleigh friction is included in the lower atmosphere to mimic the interactions with the boundary layer.

\begin{figure}
\centering
\includegraphics[width=19pc,angle=0,trim={2cm 1cm 1cm 2cm},clip]{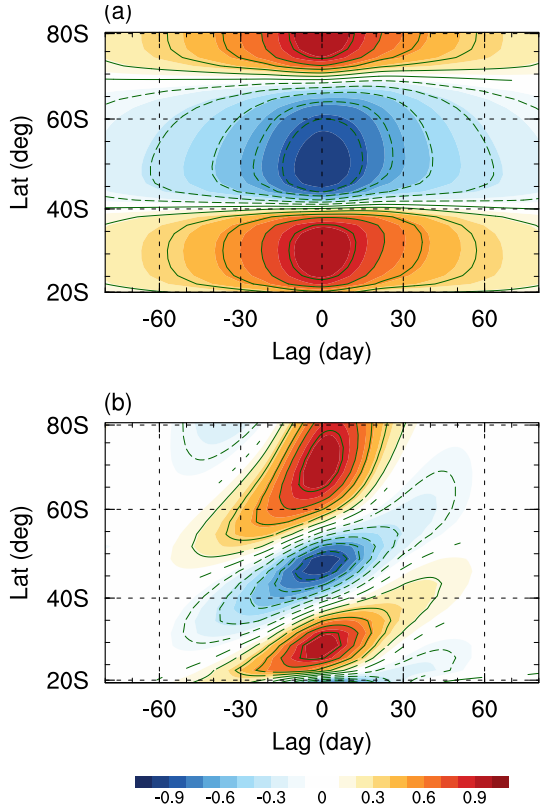}
\caption{One-point lag-correlation maps of the vertically averaged zonal-mean zonal wind anomalies $\langle\overline{u}\rangle$, reconstructed from projections onto the two leading EOFs of $\langle\overline{u}\rangle$ for the (a) non-propagating regime and (b) propagating regime in two setups of an idealized GCM. The base latitude is at 30$^{\circ}$S and the contour interval is 0.1. Regions enclosed by contour lines denote values significant at the 95\% level.}
\label{fig:1}
\end{figure}

The non-propagating and propagating regimes are produced in two slightly different setups of this model. For the setup with non-propagating regime, we use the standard configuration of \cite{Held_Suarez1994}, which employs an analytical profile approximating a troposphere in unstable radiative-convective equilibrium and an isothermal stratosphere for Newtonian relaxation. For the setup with propagating regime, we follow an approach similar to the one used by \cite{SheshadriPlumb2017}. In this setup, for the equilibrium temperature profile in the troposphere and stratosphere, we use the perpetual-solstice version of the equilibrium temperature specifications used in \cite{LubisHuang2018}, calculated from a rapid radiative transfer model (RRTM), with winter conditions in the SH. As will be seen later, these choices result in a large-scale circulation with reasonable annular mode time scales in the SH.

In Fig.~\ref{fig:1}, we show, following \citet{SonLee2006}, the one-point lag-correlation maps for the vertically averaged zonal-mean zonal wind anomalies $\langle\overline{u}\rangle$ reconstructed from projections onto the two leading EOFs of $\langle\overline{u}\rangle$ for the two setups (hereafter, angle brackets and overbars denote the vertical and zonal averages, respectively). The anomalies are defined as the deviations from the time mean.  The non-propagating and propagating regimes are clearly seen in Figs.~\ref{fig:1}a and ~\ref{fig:1}b, respectively. In the latter, the  propagating anomalies emerge in low latitudes and propagate generally poleward over the course of 3-4 months. In contrast, the non-propagating regime is characterized by persistence zonal flow anomalies in the mid-latitude (Fig.~\ref{fig:1}a).

To understand the relationship between zonal-mean zonal wind and eddy forcing in the non-propagating and propagating annular modes, the vertically averaged zonal-mean zonal wind anomalies ($\langle\overline{u}\rangle$) and vertically averaged zonal-mean eddy momentum flux convergence anomalies ($ \langle\overline{F}\rangle$) are projected onto the leading EOFs of $\langle\overline{u}\rangle$ following \citet{Lorenz2001}. The time series of  zonal index ($z$) and eddy forcing  ($m$) associated with EOF1 and EOF2 are formulated as:
\begin{equation}
z_{1,2}(t) =   \frac{ \bf{\langle\overline{u} \rangle} \mathit{(t)}\;  \bf{We_\mathrm{{1,2}}}}{\sqrt{\bf{e}^{T}_\mathrm{{1,2}}\bf{We_\mathrm{{1,2}}}}},
\end{equation}
\begin{equation}
m_{1,2}(t) =    \frac{ \bf{\langle\overline{F} \rangle} \mathit{(t)}\;  \bf{We_\mathrm{{1,2}}}}{\sqrt{\bf{e}^{T}_\mathrm{{1,2}}\bf{We_\mathrm{{1,2}}}}}, 
\end{equation}
where $z_{1,2}$ ($m_{1,2}$) denotes the component of the field $\langle\overline{u}\rangle$ ($ \langle\overline{F}\rangle$) that projects onto the latitudinal structure of the two leading EOFs. $\bf{\langle\overline{u} \rangle} \mathit{(t)}$ and $\bf{\langle\overline{F} \rangle} \mathit{(t)}$ are $\langle \overline{u} \rangle (\phi,t)$ and $\langle \overline{F} \rangle (\phi,t)$ with their latitude dimension vectorized, $\bf{W}$ is a diagonal matrix whose elements are the $\cos(\phi)$ weighting used when defining the EOF structure $\bf{e}$, and $\phi$ is latitude \citep{SimpsonShepherd2013,MaHassanzadehKuang2017}. Here, the vertically averaged zonal-mean eddy momentum flux convergence $ \langle\overline{F}\rangle$ is calculated in the spherical coordinate as:
\begin{equation}
 \langle\overline{F} \rangle (\phi,t) =- \frac{1}{\cos^2\phi} \frac{\partial (\langle\overline{u'v'}\cos^2 \phi \rangle )}{a\partial \phi }  
\end{equation}
where $u'$ and $v'$ are deviations of zonal wind and meridional wind from their respective zonal means, and $a$ is Earth's radius.

\begin{figure}
\centering
\includegraphics[width=18.5pc,angle=0,trim={3cm 4.5cm 3cm 5.5cm},clip]{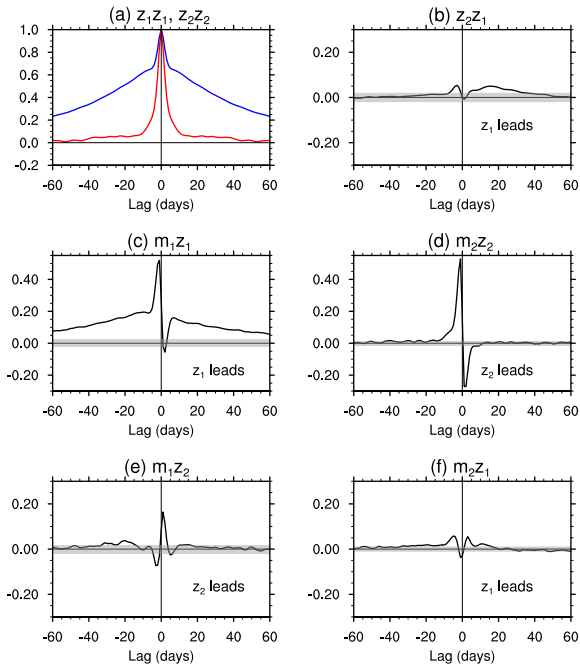}
\caption{Lagged-correlation analysis of the GCM setup with non-propagating regime. (a) Auto-correlation of $z_1$ (blue) and  $z_2$ (red), (b)
cross-correlation $z_1z_2$,  (c) cross-correlation $m_1z_1$, (d) cross-correlation $m_2z_2$, (e) cross-correlation $m_1z_2$, and (f) cross-correlation $m_2z_1$. The two leading EOFs contribute 60.2\% and 19.2\%, respectively, to the total variance. The $e$-folding decorrelation time scales of $z_1$ and $z_2$ are $64.5$ and $4.8$ days, respectively. Grey shading represents 5\% significance level according to the test of Bartlett (Appendix A).}
\label{fig:2}
\end{figure}

\begin{figure}
\centering
\includegraphics[width=18.5pc,angle=0,trim={1cm 10cm 1cm 1cm},clip]{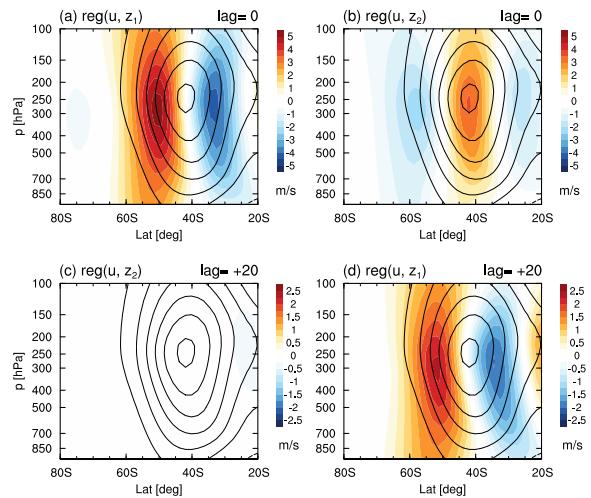}
\caption{Anomalous zonal-mean zonal wind ($\bar{u}$) regressed onto $z_1$ and $z_2$ in the GCM setup with non-propagating regime: (a,
b) simultaneous, (c) $z_2$ leads by 20 days, and (d) $z_1$ leads by 20 days. The contours are the climatological zonal-mean zonal wind with interval of 5 ms$^{-1}$.}
\label{fig:3}
\end{figure}

Figure~\ref{fig:2} shows lagged-correlation analysis between $z$ and $m$ in the GCM setup with non-propagating regime. The auto-correlation of $z_1$, as discussed in past studies \citep[e.g.,][]{ChenPlumb2009,MaHassanzadehKuang2017}, has a noticeable shoulder at around 5-day lags and shows an unrealistically persistence annular mode, well separated from the faster decaying $z_2$, which is consistent with the considerable difference in the contribution of the two EOFs to the total variance (60.2\% versus 19.2\%). The $e$-folding decorrelation time scales of $z_1$ and $z_2$ are $64.5$ and $4.8$ days, respectively.  The strong, positive cross-correlations of $m_1z_1$ and insignificant cross-correlations of $m_2z_2$ at large positive lags suggest the existence of a positive eddy-zonal flow feedback for EOF1 (from EOF1) but not for EOF2 (see \citet{SonLee2008} and \citet{MaHassanzadehKuang2017}). Figure~\ref{fig:2}b shows that the $z_1z_2$ cross-correlations are weak at positive and negative lags, which consistently with the one-point lag-correlation map of Fig.~\ref{fig:1}a  and Fig.~\ref{fig:3} (shown later), are indicative of a non-propagating regime, as reported previously for a similar setup \citep{SonLee2006,SonLee2008}. The $m_1z_2$  and $m_2z_1$ cross-correlations are small and often insignificant, suggesting the absence of the cross-EOF feedbacks in the non-propagating regime (Figs.~\ref{fig:2}e-f). All together, the above analysis shows that for the non-propagating regime, single-EOF reduced-order models such as LH01 are sufficient.  

The weak cross-correlations between $z_1$ and $z_2$ in the GCM with non-propagating regime  (Fig.~\ref{fig:2}b) can be also seen by regressing the zonal-mean zonal wind anomalies on the zonal index at 0- and 20-day time lag. Figures~\ref{fig:3}a and \ref{fig:3}b show the wind anomalies regressed on $z_1$ and $z_2$ at lag 0, yielding approximately the EOF1 and EOF2 patterns, respectively. Twenty days after $z_1$ leads zonal wind anomalies, the anomalies do not drift poleward or decay, but rather persist (Fig.~\ref{fig:3}d). In contrast, 20 days after $z_2$ leads zonal wind anomalies, the anomalies decay and disappear (Fig.~\ref{fig:3}c). These observations are consistent with the long and short persistence of $z_1$ and $z_2$, respectively, consistent with the weak cross-correlations of $z_1$ and $z_2$ at positive or negative lags, and as become clear below, consistent with the non-propagating nature of this setup. 

\begin{figure}
\centering
\includegraphics[width=18.5pc,angle=0,trim={3cm 4.5cm 3cm 5.7cm},clip]{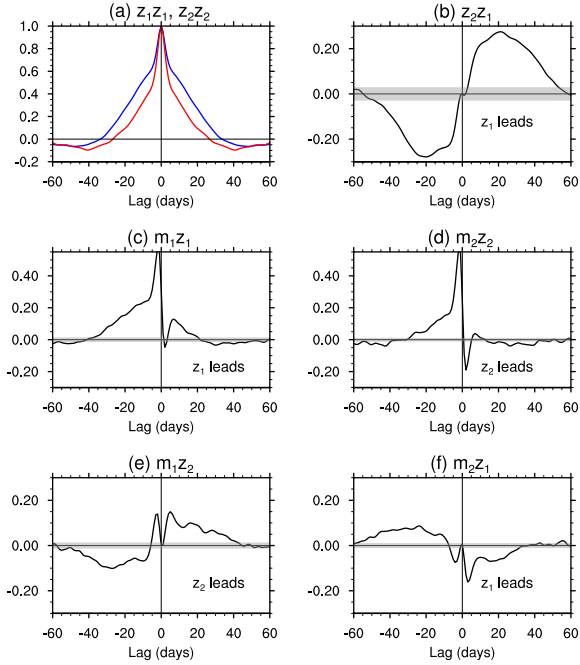}
\caption{ Lagged-correlation analysis of the GCM setup with propagating regime. (a) Auto-correlation of $z_1$ (blue) and  $z_2$ (red), (b)
cross-correlation $z_1z_2$,  (c) cross-correlation $m_1z_1$, (d) cross-correlation $m_2z_2$, (e) cross-correlation $m_1z_2$, and (f) cross-correlation $m_2z_1$. The two leading EOFs contribute 40.4\% and 32.5\%, respectively, to the total variance. The $e$-folding decorrelation time scales of $z_1$ and $z_2$ are $14.1$ and $9.2$ days, respectively.  Grey shading represents 5\% significance level according to the test of Bartlett (Appendix A).}
\label{fig:4}
\end{figure}

Figure~\ref{fig:4} shows lagged-correlation analysis between $z$ and $m$ in the GCM setup with propagating regime. The auto-correlation of $z_1$, its persistence compared to that of $z_2$, and the explained variance by the two EOFs (40.4\% versus 32.5\%) are much more similar to what is observed in the SH (shown later in Fig.~\ref{fig:7}). The $e$-folding decorrelation time scales of $z_1$ and $z_2$ are $14.1$ and $9.2$ days, respectively. Figure~\ref{fig:4}b shows that $z_1$ and $z_2$ are strongly correlated at long lags peaking at around $\pm 20$ days. This behavior along with the one-point lag-correlation map of Fig.~\ref{fig:1}b and regression map of wind anomalies (Fig. 5, shown later) suggests the existence of a propagating regime, as noted by few previous studies (e.g., \citealt{SonLee2006,SheshadriPlumb2017}). It should be noted that \citet{SonLee2006} have proposed a rule of thumb based on the ratio of the explained variance of EOF2 to EOF1: A non-propagating (propagating) regimes exists if the ratio is smaller (larger) than 0.5. The regime of our two setups are consistent with this rule of thumb as the ratios are $\sim 0.3$ and $\sim 0.8$ in our non-propagating and propagating regimes.

Furthermore, Fig.~\ref{fig:4}c shows that the $m_1z_1$ cross-correlations are positive at long positive lags (5-20 days) and then negative but small. Fig.~\ref{fig:4}d indicates small and negative cross-correlations between $z_2$ and $m_2$ at the times scale of longer than 20 days (Fig.~\ref{fig:4}c). Overall, the shape of the $m_1z_1$ and  $m_2z_2$ cross-correlation functions are similar between the non-propagating and propagating regimes, although the $m_1z_1$ cross-correlations are larger and more persistent in the non-propagating regime. In contrast, the $m_1z_2$ and $m_2z_1$ cross-correlations are substantially different between the two regimes (Figs.~\ref{fig:4}e-f). There are statistically significant and large positive $m_1z_2$ cross-correlations at large positive lags ($>$ 5~days) and statistically significant and large negative $m_2z_1$ cross-correlations at positive lags up to 30~days. Note that as emphasized in the figures, positive lags here mean that $z_1$ ($z_2$) is leading $m_2$ ($m_1$). Therefore, these cross-correlations, as discussed later, indicate the existence of cross-EOF feedbacks in the propagating regime.

\begin{figure}
\centering
\includegraphics[width=18.5pc,angle=0,trim={1cm 10cm 1cm 1cm},clip]{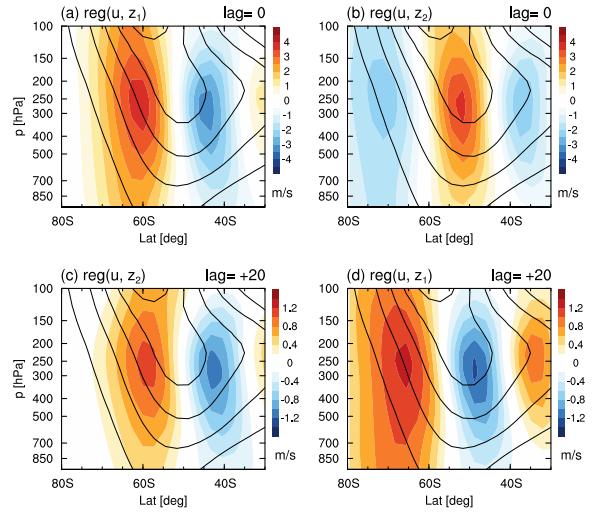}
\caption{Anomalous zonal-mean zonal wind ($\bar{u}$) regressed onto $z_1$ and $z_2$ in the GCM setup with propagating regime: (a,
b) simultaneous, (c) $z_2$ leads by 20 days, and (d) $z_1$ leads by 20 days. The contours are the climatological zonal-mean zonal wind with interval of 5 ms$^{-1}$.}
\label{fig:5}
\end{figure}

Figure~\ref{fig:5} shows anomalous zonal-mean zonal wind regressed on $z_1$ and $z_2$ at 0- and 20-day time lag in the GCM setup with propagating regime. Figures~\ref{fig:5}a and \ref{fig:5}b show the wind anomalies regressed on $z_1$ and $z_2$ at lag 0, again yielding approximately the EOF1 and EOF2 patterns, respectively. As shown in Fig.~\ref{fig:5}c, 20 days after $z_2$ leads zonal wind anomalies, the anomalies have drifted poleward and project strongly onto the structure of wind anomalies associated with EOF1 (Figs.~\ref{fig:5}a,c, pattern correlation = 0.93). This is consistent with positive correlation of $z_1z_2$ at lag +20 days when $z_1$ leads $z_2$ (Fig.~\ref{fig:4}b). Likewise, twenty days after $z_1$ leads zonal wind anomalies, the anomalies (of Fig.~\ref{fig:5}a) have drifted poleward and project strongly onto the structure of anomalies associated with EOF2, but with an opposite sign (Figs.~\ref{fig:5}b,d, pattern correlation = -0.85). This is consistent with negative correlation of $z_1z_2$ when $z_2$ leads $z_1$ by 20 days (Fig.~\ref{fig:4}b).

Overall, these results suggest the existence of cross-EOF feedbacks in the propagating annular mode. In Section~3, we will developed a model to quantify these four feedbacks and understand the effects of their magnitude and signs on the variability (e.g., persistence) of $z_1$ and $z_2$. But first, we will examine the variability and characteristics of $z$ and $m$ in reanalysis. In particular, we will see that the $z$ and $m$ cross-correlations in the GCM's propagating regime well resemble those in the SH reanalysis data.

\subsection{Reanalysis}
\label{sec:22}

We use the 1979-2013 data from the European Centre for Medium-Range Weather Forecasts (ECMWF) interim reanalysis (ERA-Interim; \citealt{Dee2011}). Zonal and meridional wind components $(u,v)$ are 6 hourly, on 1.5$^{\circ}$ latitude $\times$ 1.5$^{\circ}$ longitude grid, and on 21 vertical levels between 1000 and 100 hPa. Anomalies used for computing correlations and EOF analyses are defined as the deviations from the climatological seasonal cycle. The mean seasonal cycle is defined as the annual average and the first four Fourier harmonics of the 35-yr daily climatology.

\begin{figure}
\centering
\includegraphics[width=19pc,angle=0,trim={4cm 8.5cm 3cm 1cm},clip]{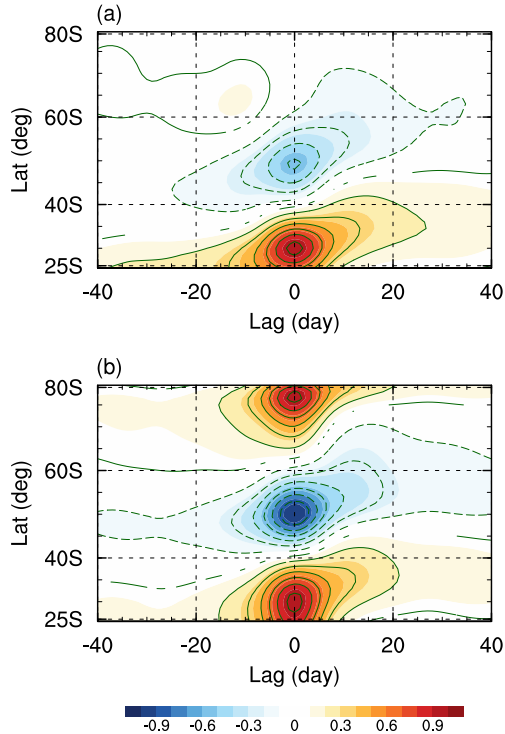}
\caption{One-point lag-correlation maps of the vertically averaged zonal-mean zonal wind anomalies from year-round ERA-Interim data integrated across the depth of the troposphere (1000-100 hPa) ($\langle\overline{u}\rangle$) in the Southern Hemisphere. (a) Total anomaly fields and (b) reconstructed from projections onto the two leading EOFs of $\langle\overline{u}\rangle$. The base latitude is at 30$^{\circ}$S and the contour interval is 0.1. Regions enclosed by contour lines denote values significant at the 95\% level according to the $t$-test.}
\label{fig:6}
\end{figure}

Figure~\ref{fig:6} shows a one-point lag-correlation map of vertically averaged zonal-mean zonal wind $\langle\overline{u}\rangle$ in the SH, where the base latitude is 30$^{\circ}$S. Comparing this figure with Fig.~\ref{fig:1}, it can be seen that there is an indication of poleward-propagating anomalies in SH, which appear in low latitudes and migrate poleward over the course of 2-3 months (Fig.~\ref{fig:6}a). However, the poleward-propagating signals are not as clearly as those observed in the GCM setup with the propagating regime (Fig.~\ref{fig:1}b, or Fig. 2 of \citealt{SonLee2006}). This is consistent with previous studies (e.g. \citealt{Feldstein1998,FeldsteinLee1998,SheshadriPlumb2017}), showing that both propagating and non-propagating anomalies exist in all seasons in the SH, which somehow obscure the propagating signals. Reconstructions based on the projections onto the two leading EOFs of zonal-mean zonal wind further show that most of the mid-latitude SH wind variability can be explained by the two leading EOF modes (Fig.~\ref{fig:6}b). The ratio of the fractional variance of EOF2 (23.2\%) to that of EOF1 (45.1\%) is 0.51, which is right at the boundary from the rule of thumb. Overall, as already pointed out by \cite{SheshadriPlumb2017}, a propagating annular mode exists in the SH and is largely explained by the two leading EOF modes.

\begin{figure}
\centering
\includegraphics[width=19pc,angle=0,trim={3cm 4.5cm 3cm 5cm},clip]{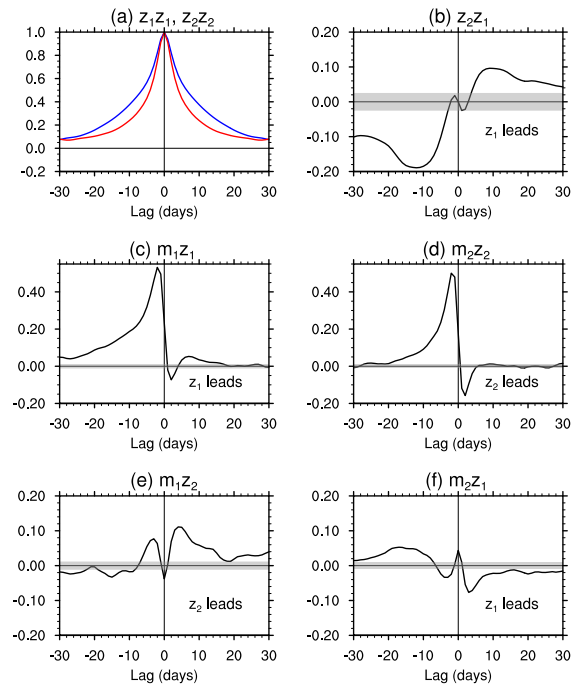}
\caption{Lagged-correlation analysis for the Southern Hemisphere, calculated from year-round ERA-Interim data. (a) Auto-correlations of $z_1$ (blue) and $z_2$ (red), (b) cross-correlation $z_1z_2$,  (c) cross-correlation $m_1z_1$, (d) cross-correlation $m_2z_2$, (e) cross-correlation $m_1z_2$, and (f) cross-correlation $m_2z_1$ at different lags. The two leading EOFs contribute to 45.1\% and 23.2\% of the total variance, respectively. The $e$-folding decorrelation time scales of $z_1$ and $z_2$ are $10.3$ and $8.1$ days, respectively. Grey shading represents 5\% significance level according to the test of Bartlett (Appendix A).}
\label{fig:7}
\end{figure}

Figure~\ref{fig:7}a shows the auto-correlations of $z_1$ and $z_2$.  Consistent with \citet{Lorenz2001}, the estimated decorrelation time scales of these two PCs are 10.3 and 8.1 days, respectively. Figure~\ref{fig:7}b depicts the cross-correlation $z_1z_2$, showing statistically significant and relatively strong correlations that peak around $\pm 10$~days. As discussed in earlier studies, such lagged correlations are a signature of the propagating annular modes \citep{FeldsteinLee1998,SonLee2006,SonLee2008,SheshadriPlumb2017}, implying that the period of the poleward propagation is about 20-30 days in the SH (Fig. ~\ref{fig:7}b), consistent with \cite{SheshadriPlumb2017} and with Fig.~\ref{fig:6}.

To understand the effects of $z_1$ and $z_2$ on $m_1$ and $m_2$, we also examine the cross-correlations between $z$ and $m$ at different lags (Figs.~\ref{fig:7}c-f). The shape and the magnitude of the $m_1z_1$ and $m_2z_2$ cross-correlations (Figs.~\ref{fig:7}c-d) are similar to those originally shown by \citet{Lorenz2001} (see their Figs.~5 and 13a) and later by many others using different reanalysis products and time periods. As discussed in \citet{Lorenz2001}, the statistically significant positive $m_1z_1$ cross-correlations at long positive lags ($\sim 8-20$~days) and the insignificant $m_2z_2$ cross-correlations for time scales longer than $\sim 5$~days are indicative that a positive eddy-zonal flow feedback exists only for EOF1, but not for EOF2 (also see \citet{Byrne2016} and \citet{MaHassanzadehKuang2017}). We emphasize that this positive feedback is from EOF1 onto itself.   

To see if there are cross-EOF feedbacks, in Figs.~\ref{fig:7}e-f we plot the $m_1z_2$ and $m_2z_1$ cross-correlations at different lags. The $m_1z_2$ cross-correlations show statistical significant positive correlations at large positive lags, signifying that a cross-EOF feedback, i.e, $z_2$ modifying $m_1$, is present. Note that the magnitude of the $m_1z_2$ cross-correlations at positive lags is overall larger than those of $m_1z_1$ (Fig.~\ref{fig:7}c). There are also statistically significant but negative $m_2z_1$ correlations at large positive lags, again suggesting the existence of a cross-EOF feedback, i.e, $z_1$ modifying $m_2$. These results indicate that in the presence of propagating regime in the SH, there are indeed cross-EOF feedbacks; however, these feedbacks were always ignored in the previous studies and reduced-order models of the SH extratropical large-scale circulation.

\section{Eddy-zonal flow feedbacks in the propagating annular modes: Model and quantification}
\label{sec:3}

In this section, an eddy-zonal flow feedback model that accounts for the coupling of the leading two EOFs and their feedbacks, including the cross-EOF feedbacks will be introduced. Then this model will be validated using synthetic data from a simple stochastic prototype, and from its analytical solution, we will derive conditions for the existence of the propagating regime. Finally, we will use this model to estimate the feedback strengths of the propagating annular modes in data from the reanalysis (SH) and the idealized GCM.

\subsection{Developing an eddy-zonal flow feedback model for propagating annular modes}
\label{sec:31}

With the same notations as in \citet{Lorenz2001}, the time series of zonal indices ($z_{1}$ and $z_{2}$) and eddy forcing ($m_{1}$ and $m_{2}$) associated with the first two leading EOFs are calculated by projecting the vertically averaged zonal-mean zonal wind $\langle\overline{u}\rangle$ and eddy momentum flux convergence $\langle\overline{F}\rangle$ anomalies onto the patterns of the first and second EOFs of $\langle\overline{u}\rangle$ (see Eqs.~(1)-(2)). Equations for the tendency of $z_1$ and $z_2$ can be then formulated as:

\begin{equation}
\frac{dz_1}{dt}=m_1 -\frac{z_1}{\tau_1}, \label{eq:z1}
\end{equation}
\begin{equation}
\frac{dz_2}{dt}=m_2 -\frac{z_2}{\tau_2}, \label{eq:z2}
\end{equation}
where $t$ is time and the last term on the right-hand side in each equation represents damping (mainly due to surface friction) with time scale $\tau$. As discussed in \citet{Lorenz2001}, Eqs.~(4)-(5) can be interpreted as the zonally and vertically averaged zonal momentum equation:
\begin{equation}
\frac{\partial \langle\overline{u}\rangle }{\partial t } = - \frac{1}{\cos^2\phi} \frac{\partial (\langle\overline{u'v'}\cos^2 \phi \rangle )}{a\partial \phi } - D ,
\end{equation}
projected into EOF1 and EOF2, respectively. In the above equation, $D$ includes the effects of surface drag and is modeled as Rayleigh drag in Eqs.~(4)-(5).

Assuming a linear representation for the feedback of an EOF onto itself, \citet{Lorenz2001} and later studies wrote $m_1(t)=\tilde{m}_1(t)+b_1 z_1(t)$ and $m_2(t)=\tilde{m}_2(t)+b_2 z_2(t)$, where $b_1$ and $b_2$ are the feedback strengths (with $b_j>0$ implying a positive feedback that prolongs the persistence of $z_j$). $\tilde{m}$ is the random, zonal flow-independent component of the eddy forcing that drives the high-frequency variability of $z$ \citep{Lorenz2001,MaHassanzadehKuang2017}.

Here, to account for the cross-EOF feedbacks, i.e., the effect of $z_2$ on $m_1$ and $z_1$ on $m_2$, we extend the LH01 model and write 

\begin{equation}
m_1=\tilde{m}_1+b_{11}z_1+b_{12}z_2,\label{eq:m1}
\end{equation}

\begin{equation}
m_2=\tilde{m}_2+b_{21}z_1+b_{22}z_2. \label{eq:m2}
\end{equation}

With $j,k=1,2$, $b_{jk}$ is the strength of the linearized feedback of $z_k$ onto $z_j$ through modifying $m_j$ in the quasi-steady limit; thus the cross-EOF feedbacks are represented by the terms involving $b_{12}$ and $b_{21}$. To find the values of $b_{jk}$, we can use the lagged-regression method of \citet{SimpsonShepherd2013}, which assumes that $reg_l(\tilde{m}_j,z_j)=sum(\tilde{m}_j(t+l)z_j(t)) \approx 0$ at large positive lags, $l$. By lag-regressing each term in Eqs.~(\ref{eq:m1}) onto $z_1$ and then onto $z_2$, we find
\begin{equation}
\begin{bmatrix}
reg_l(z_1, z_1) & reg_l(z_2, z_1) \\ 
reg_l(z_1, z_2) & reg_l(z_2, z_2)
\end{bmatrix} 
\begin{bmatrix}
b_{11}\\ 
b_{12}
\end{bmatrix}
=
\begin{bmatrix}
reg_l(m_1, z_1)\\ 
reg_l(m_1, z_2)
\end{bmatrix} \label{eq:b11}
\end{equation}
and similarly, from Eq.~(\ref{eq:m2}) we find
\begin{equation}
\begin{bmatrix}
reg_l(z_2, z_1) & reg_l(z_1, z_1) \\ 
reg_l(z_2, z_2) & reg_l(z_1, z_2)
\end{bmatrix} 
\begin{bmatrix}
b_{21}\\ 
b_{22}
\end{bmatrix}
=
\begin{bmatrix}
reg_l(m_2, z_1)\\ 
reg_l(m_2, z_2)
\end{bmatrix}, \label{eq:b22}
\end{equation}
where we assumed $reg_l(\tilde{m}_j,z_k) \approx 0$ for $j,k=1,2$. 

\begin{table}
\caption{Prescribed and estimated feedback strengths (in day$^{-1}$) in synthetic data for the case without cross EOF-feedbacks. The imposed damping rates of friction are $\tau_1$=$\tau_2$= $8$~days. The values of $b$ and $\tau$ are motivated by the observed ones, see Table~\ref{tab:4}.}
\begin{center}
\begin{tabular}{ccccccrrcrcrcrc}
\topline
Feedback                   & $b_{11} $    & $b_{12}$  	& $b_{21}$ 	& $b_{22}$ \\
\midline
Prescribed                        & 0.040             & 0.000 		& 0.000			& 0.000 \\
Estimated (Eqs.~(\ref{eq:b11})-(\ref{eq:b22}))   & 0.042             & 0.001  	& -0.0006			& 0.0005\\
\botline \label{tab:1}
\end{tabular}
\end{center}
\end{table}

Note that if one attempts to find $b_{11}$ using a single-EOF approach such as LH01, then, from Eq.~(\ref{eq:m1}), one would be implicitly assuming that 
$reg_l(\tilde{m}_1+b_{12}z_2,z_1)=reg_l(\tilde{m}_1,z_1)+b_{12} reg_l(z_2,z_1) \approx b_{12} reg_l(z_2,z_1)$ is zero. However, as shown earlier, in the propagating regime, the $z_1z_2$ cross-correlations can be large at long lags, and as discussed below, the range of time lags needed to be used in Eqs.~(\ref{eq:b11})-(\ref{eq:b22}) and the lags at which $z_1z_2$ cross-correlations peaks are often comparable. Consequently, if $b_{12} \neq 0$, the key assumption of the statistical methods developed to quantify eddy-zonal flow feedbacks \citet{Lorenz2001,SimpsonShepherd2013,MaHassanzadehKuang2017}) is violated. Therefore, $b_{jk}$ should be determined together by solving the systems of equations (\ref{eq:b11})-(\ref{eq:b22}).      

The basic assumptions of our model, Eqs.~(\ref{eq:z1})-(\ref{eq:b22}), are similar to those of the LH01 model: i) A linear representation of the feedbacks is sufficient, and ii) The eddy forcing $m$ does not have long-term memory independent of the variability in the jet (represented by $z$). The second assumption means that at sufficiently large positive lags (beyond the time scales over which there is significant auto-correlation in $\tilde{m}$) the feedback component of the eddy forcing will dominate the ${m_jz_k}$ cross-correlations \citet{Lorenz2001,ChenPlumb2009,SimpsonShepherd2013,MaHassanzadehKuang2017}), i.e., $reg_l(\tilde{m}_j,z_k) \approx 0$ at  ``large-enough'' positive lags. Note that one cannot use a lag that is too long because then even $reg_l(z_j,z_j)$ would be small and inaccurate. To find the appropriate lag to use, one must look for non-zero $m_jz_k$ cross-correlations at positive lags beyond an eddy lifetime. In this study, the strengths of the individual feedbacks are averaged over positive lags of 8 to 20 days for both GCM and reanalysis (e.g., \citealt{SimpsonShepherd2013,Burrows2016}). We choose this range in order to avoid the high-frequency variability at short lags (indicated by impulsive and oscillatory characters of the $\tilde{m}$ auto-correlation) and strong damping at the very long lags.

In the following section, we will present a proof of concept for this eddy-zonal flow feedback model using synthetic data obtained from a simple stochastic prototype and show that using Eqs.~(\ref{eq:b11})-(\ref{eq:b22}), the prescribed feedbacks can be accurately backed out.

\subsection{Validation using synthetic data from a simple stochastic prototype}
\label{sec:32}

We begin by constructing a simple stochastic system to produce synthetic time series $z$ and $m$ in the presence or absence of cross-EOF feedbacks. The equations of this system are the same as Eqs.~(\ref{eq:z1})-(\ref{eq:z2}) and (\ref{eq:m1})-(\ref{eq:m2}). Following \citet{SimpsonShepherd2013}, we generate a synthetic time series of the random component of the eddy forcing $\widetilde{m}_{1,2}$ using a second-order autoregressive (AR2) noise process:

\begin{equation}
\tilde{m}_{1} (t)=0.6\tilde{m}_{1} (t-2)-0.3\tilde{m}_{1} (t-1)+\epsilon_{1} (t),  \label{eq:stoch1}
\end{equation}

\begin{equation}
\tilde{m}_{2}(t)=0.6\tilde{m}_{2}({t-2})-0.3\tilde{m}_{2}(t-1)+\epsilon_{2}(t), \label{eq:stoch2}
\end{equation}
where $t$ denotes time (in days) and $\epsilon$ is white noise distributed uniformly between -1 and +1. 

Synthetic time series of $z_{j}$ and $m_{j}$ are produced by numerically integrating Eqs.~(\ref{eq:z1})-(\ref{eq:z2}), (\ref{eq:m1})-(\ref{eq:m2}), and (\ref{eq:stoch1})-(\ref{eq:stoch2}) forward in time with two different sets of prescribed $b_{jk}$. In the first set, there is no cross-EOF feedback, i.e., $b_{12}=b_{21}=0$ (Table~\ref{tab:1}). In the second set, $b_{11}$ and $b_{22}=0$ are the same as the first set, but here there is cross-EOF feedback, i.e., $b_{12}$ and $b_{21} \neq 0$ (Table~\ref{tab:2}). For both sets, we assumed $\tau_1=\tau_2= 8 $ days. The values of $b$ and $\tau$ are reasonably chosen based on the observed values in the SH (see Table 4).

\begin{table}
\caption{Prescribed and estimated feedback strengths (in day$^{-1}$) in synthetic data for the case with cross EOF-feedbacks. The imposed damping rates of friction are $\tau_1$=$\tau_2$= $8$~days. The values of $b$ and $\tau$ are motivated by the observed ones, see Table~\ref{tab:4}.}
\begin{center}
\begin{tabular}{ccccccrrcrcrcrc}
\topline
Feedback                   & $b_{11} $    & $b_{12}$  	& $b_{21}$ 	& $b_{22}$ \\
\midline
Prescribed                      & 0.040             & 0.060 		& -0.025			& 0.000\\
Estimated (Eqs.~(\ref{eq:b11})-(\ref{eq:b22}))   & 0.043             & 0.067 		& -0.026			& -0.002\\
\botline \label{tab:2}
\end{tabular}
\end{center}
\end{table}

\begin{figure}
\centering
\includegraphics[width=18.5pc,angle=0,trim={0cm 0cm 0cm 0cm},clip]{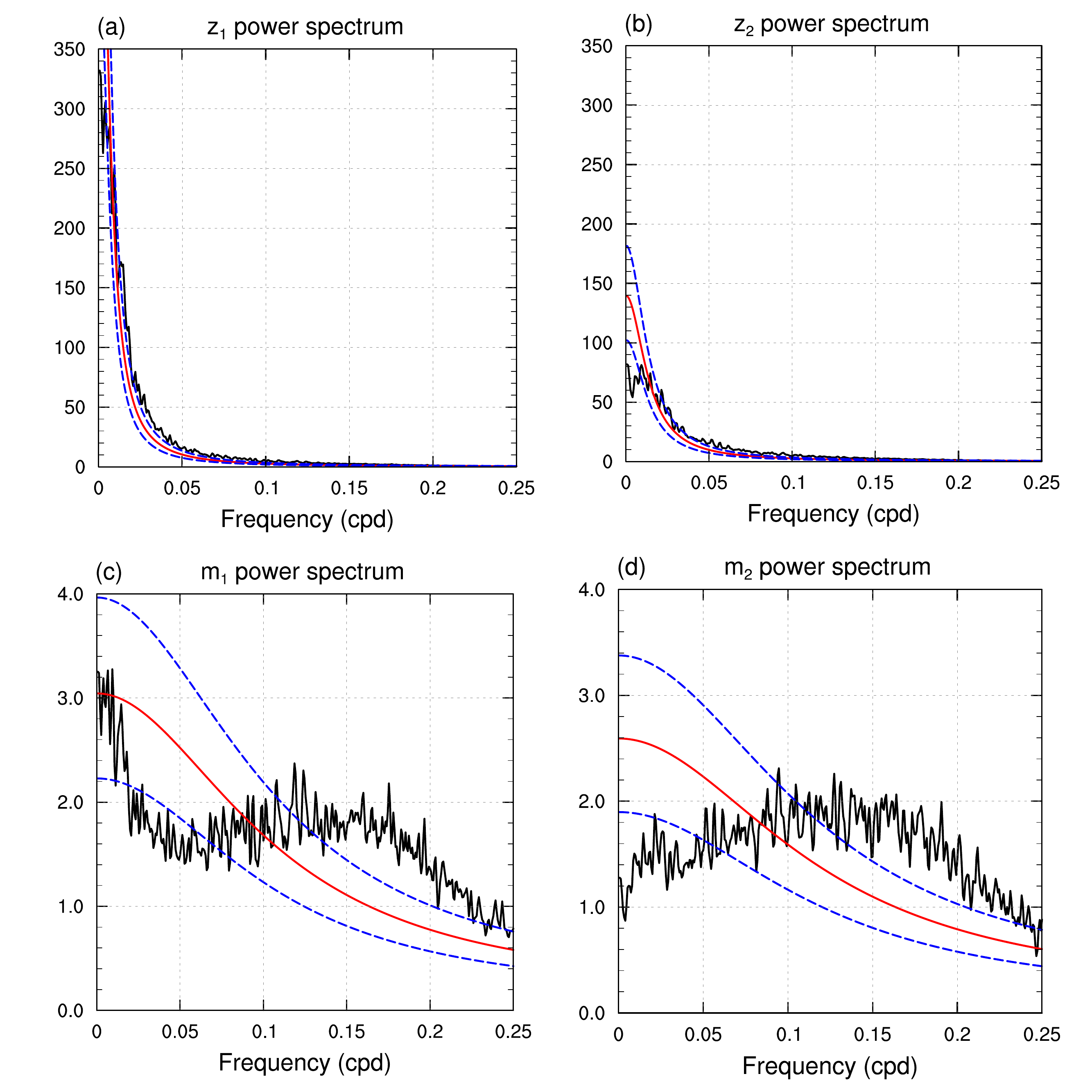}
\caption{Spectra of $z_{1,2}$ and $m_{1,2}$ from the synthetic data with cross-EOF feedbacks. Black lines show the power spectra of (a) $z_1$, (b) $z_2$, (c) $m_1$, and (d) $m_2$. The red-noise spectra are indicated by the smooth solid red curves, and the smooth dashed blue lines are the 5\% and 95\% a priori confidence limits.}
\label{fig:8}
\end{figure}

Spectral analysis of $z_{1,2}$ and $m_{1,2}$ shows that the synthetic data indeed have characteristics similar to those of the observed SH. For example, for the case with cross-feedbacks (Fig.~\ref{fig:8}), we find that consistent with observations (see Fig.~4 of \citet{Lorenz2001} or Fig.~3 of \citet{MaHassanzadehKuang2017}), the time scales of $z_1$ and $z_2$ are much longer (i.e., slower variability) than $m_1$ and $m_2$, and the power spectra of $z$ can be interpreted, to the first order, as reddening of the power spectra of eddy forcing $m$ \citet{Lorenz2001,MaHassanzadehKuang2017}.  The power spectra of eddy forcings $m_1$ and $m_2$ have in general a broad maximum centered at the low and synoptic frequency, consistent with observations. Given that the characteristics of the synthetic data mimic the key characteristics of the observed annular modes, we use this idealized framework to validate the lagged-correlation approach of Eqs.~(\ref{eq:b11})-(\ref{eq:b22}) for quantifying eddy-zonal flow feedbacks.

\begin{figure}
\centering
\includegraphics[width=18.5pc,angle=0,trim={3cm 4.4cm 3cm 5.7cm},clip]{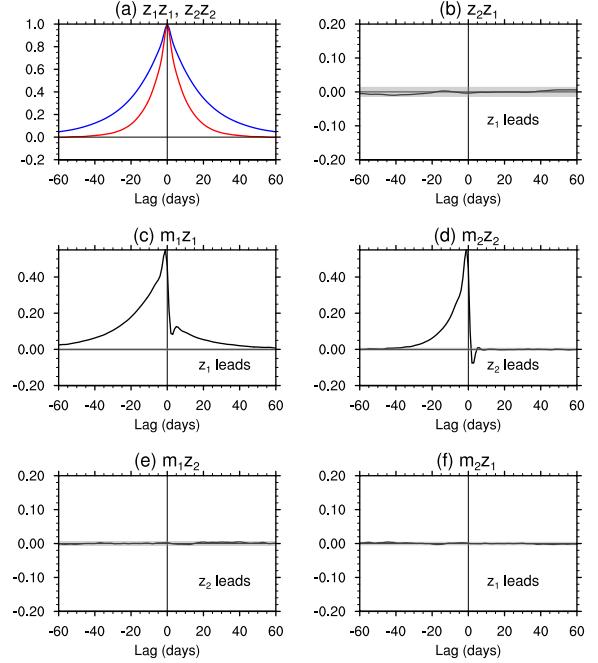}
\caption{Lagged-correlation analysis of synthetic data without cross-EOF feedback. (a) Auto-correlation of $z_1$ (blue) and  $z_2$ (red), (b)
cross-correlation $z_1z_2$,  (c) cross-correlation $m_1z_1$, (d) cross-correlation $m_2z_2$, (e) cross-correlation $m_1z_2$, and (f) cross-correlation $m_2z_1$. The $e$-folding decorrelation time scales of $z_1$ and $z_2$ are $18.6$ and $9.2$ days, respectively. Grey shading represents 5\% significance level according to the test of Bartlett (Appendix A).}
 \label{fig:9}
\end{figure}

Figure~\ref{fig:9} shows the lagged-correlation analysis of the synthetic data without cross-EOF feedbacks. It is clearly seen that the only noticeable cross-correlations are that of $m_1z_1$, and there are no (statistically significant) cross-correlations between $z_1z_2$, $m_1z_2$ and $m_2z_1$ at any lag, consistent with a non-propagating regime and the absence of cross-EOF feedbacks (Fig.~\ref{fig:2}). Using Eqs.~(\ref{eq:b11})-(\ref{eq:b22}) and lag $l$=8-20 days, we can closely estimate the prescribed feedback parameters, i.e., $b_{11}=0.04$~day$^{-1}$  and $b_{22}=b_{12}=b_{21}=0$ (see Table~\ref{tab:1}). 
 
\begin{figure}
\centering
\includegraphics[width=18.5pc,angle=0,trim={3cm 4.4cm 3cm 5.7cm},clip]{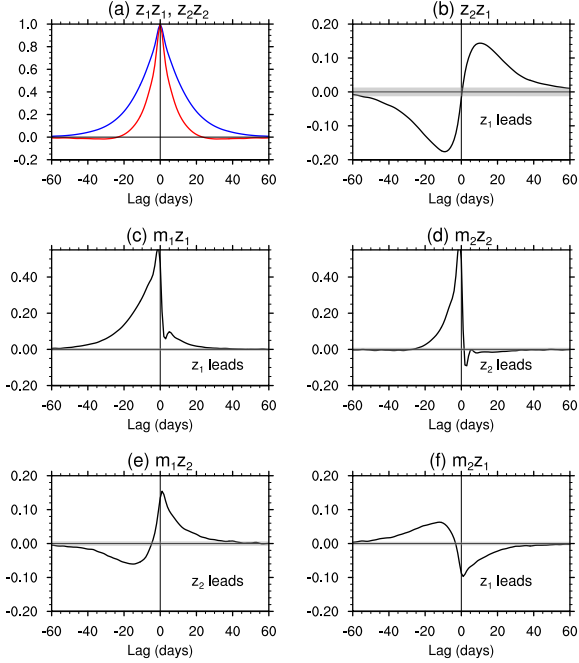}
\caption{Lagged-correlation analysis of synthetic data with cross-EOF feedback. (a) Auto-correlation of $z_1$ (blue) and  $z_2$ (red), (b)
cross-correlation $z_1z_2$,  (c) cross-correlation $m_1z_1$, (d) cross-correlation $m_2z_2$, (e) cross-correlation $m_1z_2$, and (f) cross-correlation $m_2z_1$.  The $e$-folding decorrelation time scales of $z_1$ and $z_2$ are $13.9$ and $6.5$ days, respectively. The regions outside the gray shading indicate 95\% significance level according to the test of Bartlett (Appendix A).}
 \label{fig:10} 
\end{figure}

Figure~\ref{fig:10} shows the lagged-correlation analysis of the synthetic data with cross-EOF feedbacks. First, we see that there are statistically significant and often large cross-correlations in $z_1z_2$, $m_1z_1$, $m_1z_2$, and $m_2z_1$, with the shape of the cross-correlation distributions not that different from that of the SH reanalysis and the idealized GCM setup with propagating regime (Figs.~\ref{fig:4} and \ref{fig:7}). The positive $m_1z_1$ and near zero $m_2z_2$ cross-correlations at large positive lags signify a positive $z_1$-onto-$z_1$ feedback through $m_1$, but no $z_2$-onto-$z_2$ feedback through $m_2$, consistent with the prescribed positive value of $b_{11}$ and $b_{22}=0$.  In addition, Figs.~\ref{fig:10}e-f also show that there are statistically significant and large correlations in $m_1z_2$ and $m_2z_1$ at positive lags, consistent with the introduction of cross-EOF feedbacks by setting $b_{12}=0.06$~day$^{-1}$ and $b_{21}=-0.025$~day$^{-1}$. The positive $m_1z_2$ cross-correlations are positive lags are higher than those of $m_1z_1$ (note that $b_{12}/b_{11} \approx 1.5$), and the sign of $m_2z_1$ cross-correlations is opposite to the sign of $m_1z_2$ cross-correlations (note that $b_{12}b_{21} <0$). Using Eqs.~(\ref{eq:b11})-(\ref{eq:b22}) and lag $l$=8-20 days, we can again closely estimate the prescribed feedback parameters, including the strength of the cross-EOF feedbacks (see Table~\ref{tab:2}).

The above analyses validate the approach using Eqs.~(\ref{eq:b11})-(\ref{eq:b22}) for quantifying the feedback strengths $b_{jk}$ in data from both propagating and non-propagating regimes. Furthermore, a closer examination of $z_1$ and $z_2$ auto-correlations in Figs.~\ref{fig:9}a and \ref{fig:10}a show that both $z_1$ and $z_2$ in the case without cross-EOF feedbacks are more persistence than those in the case with cross-EOF feedbacks; e.g., the $e$-forcing deccorelation time scale of $z_1$ is $18.6$~days in Fig.~\ref{fig:9}a while it is $13.9$~days in Fig.~\ref{fig:10}a. This observation might be counter-intuitive because both cases have the same $b_{11}>0$ while the case with cross-EOFs feedback has $b_{12}>0$, which might seem like another positive feedback that should further prolong the persistence of $z_1$. Finally, we notice that $b_{12}b_{21}<0$ in Table~\ref{tab:2} and in the SH reanalysis and idealized GCM setup with the propagating regime (Tables~\ref{tab:4} and \ref{tab:5}). Synthetic data generated with the same parameters as in Table~\ref{tab:2} but with the sign of $b_{21}$ flipped results in cross-correlation distributions that are vastly different from those of Fig.~\ref{fig:10} and what is seen in the SH reanalysis and idealized GCM. Inspired by these observations, next we examine the analytical solution of the deterministic version of Eqs.~(\ref{eq:z1})-(\ref{eq:z2} and (\ref{eq:m1})-(\ref{eq:m2}) to better understand the impacts of the strength and sign of $b_{jk}$ on the variability and in particular the persistence of $z_1$ and $z_2$.

\subsection{Analytical solution of the two-EOF eddy-zonal flow feedback model}
\label{sec:33}

We focus on the deterministic (i.e., $\tilde{m}_j=0$) version of Eqs.~(\ref{eq:z1})-(\ref{eq:z2}) and (\ref{eq:m1})-(\ref{eq:m2}), which can be re-written as the following system of ordinary differential equations (ODEs):

\begin{equation}
\dot{\mathbf{z}}= \mathbf{A}\mathbf{z},
\end{equation}
where

\begin{equation}
\mathbf{z}=\begin{bmatrix}
z_1 \\
z_2
\end{bmatrix},\; \; \; \;  \mathrm{and} \;  \; \; \; 
\mathbf{A}=\begin{bmatrix}
b_{11}-\frac{1}{\tau_1} & b_{12}  \\ 
b_{21} & b_{22}-\frac{1}{\tau_2}
\end{bmatrix}.
\end{equation}

The solution to this system is
\begin{equation}
\mathbf{z}(t)=e^{\mathbf{A} t} \mathbf{z}(0) = \left[\mathbf{V} e^{\boldsymbol{\Lambda} t} \mathbf{V}^{-1} \right] \mathbf{z}(0),\label{eq:sol}
\end{equation}
where $\mathbf{V}$ and $\boldsymbol{\Lambda}$ are the eigenvector and eigenvalue matrices of $\mathbf{A}$:
\begin{equation}
\mathbf{V} = \left[\mathbf{v}_1 \;\; \mathbf{v}_2 \right]
=
\begin{bmatrix}
v_{11} & v_{12} \\ 
v_{21} & v_{22}
\end{bmatrix}, \; \; \; \; 
\mathrm{and} \; \; \; \; 
\boldsymbol{\Lambda} = 
\begin{bmatrix}
\lambda_1 & 0 \\ 
0 & \lambda_2
\end{bmatrix}.
\end{equation}

To find the eigenvalues $\lambda$, we set the determinant of $\mathbf{A}$ equal to zero and solve the resulting quadratic equation to obtain: 
\begin{equation}
\begin{split}
\lambda_{1,2}=-  \frac{1}{2}  \left ( \frac{1}{\tau_1}+\frac{1}{\tau_2}-b_{11}-b_{22}\right ) \pm  \\  \frac{1}{2}  \sqrt{\left \{ \left (\frac{1}{\tau_1}-\frac{1}{\tau_2}  \right ) - \left ( b_{11}-b_{22} \right )  \right \}^2+ 4 b_{12} b_{21} }  ,
\end{split}
\end{equation} 
which, in the limit of $\tau_1 \approx  \tau_2$ (reasonable given their estimated values in Tables~\ref{tab:4} and \ref{tab:5}), simplifies to: 
\begin{equation}
\lambda_{1,2}=-  \frac{1}{2} \left ( \frac{2}{\tau}-b_{11}-b_{22}\right )  \pm \frac{1}{2} \sqrt{  \left ( b_{11}-b_{22} \right )^2 + 4 b_{12} b_{21}}.\label{eq:lambda}
\end{equation}

The solution (Eq.~(\ref{eq:sol})) can be re-written as
\begin{equation}
\mathbf{z}=c_1 e^{\lambda_1 t}\mathbf{v}_1+c_2 e^{\lambda_2 t}\mathbf{v}_2, 
\end{equation} 
where $c_1$ and $c_2$ depend on the initial condition. 

This system has a decaying-oscillatory solution, i.e., is in the propagating regime, if and only if the eigenvalues (\ref{eq:lambda}) have non-zero imaginary parts, which requires, as a necessary and sufficient condition:
\begin{equation}
\left ( b_{11}-b_{22} \right )^2  <  -4 b_{12} b_{21}. \label{eq:condition1}
\end{equation} 
Equation~(\ref{eq:condition1}) also implies that a necessary condition for the existence of propagating regimes is
\begin{equation}
b_{12} b_{21}<0. \label{eq:condition2}
\end{equation} 
Thus, non-zero cross-EOF feedbacks of opposite signs are essential components of the propagating regime dynamics. The propagating regimes in the stochastic prototype (Table~\ref{tab:2}), SH reanalysis (Table~\ref{tab:4}), and idealized GCM (Table~\ref{tab:5}) satisfy the conditions of Eqs.~(\ref{eq:condition1})-(\ref{eq:condition2}), while the non-propagating regimes (Tables~\ref{tab:1} and \ref{tab:5}) do not.

In the non-propagating regime, $\lambda_{1,2}=-\sigma_{1,2} < 0$ and $\mathbf{v}_{1,2}$ are real and in this regime, $z_{1,2}$ just decay exponentially according to 
\begin{equation}
\mathbf{z}=c_1 e^{(-\sigma_1 t)}\mathbf{v}_1+c_2 e^{(- \sigma_2 t)}\mathbf{v}_2.
\end{equation} 

In the propagating regime, $\lambda_{1,2}=-\sigma \pm i \omega $ and $\mathbf{v}_{1,2}$ are complex where
\begin{eqnarray}
\sigma &=&  \frac{1}{2}  \left ( \frac{1}{\tau_1}+\frac{1}{\tau_2}-b_{11}-b_{22}\right ), \label{eq:sigma}\\
\omega &=& \frac{1}{2}  \sqrt{\left \{ \left (\frac{1}{\tau_1}-\frac{1}{\tau_2}  \right ) - \left ( b_{11}-b_{22} \right )  \right \}^2+ 4 b_{12} b_{21} }. \label{eq:omega}
\end{eqnarray}
In this regime, $z_{1,2}$ decay and oscillate according to
\begin{equation}
\mathbf{z}=c_1 e^{(-\sigma t)}e^{(i \omega t)} \mathbf{v}_1+c_2 e^{(-\sigma t)}e^{(- i \omega t)}\mathbf{v}_2. 
\end{equation} 
Realizing that in this case $v_{11}=v_{12}$ are real, and $v_{21}=v^*_{22}$ and $c_{1}=c^*_{2}=c$, where $^*$ means complex conjugate, we can re-write the above equations as
\begin{eqnarray}
z_1&=&\left[c \, e^{( i \omega t)}  v_{11}  + c^* e^{(- i \omega t)} v_{11}  \right]  e^{(-\sigma t)} , \label{eq:z1SOL}\\
z_2&=&\left[c \, e^{( i \omega t)}  v^*_{22}+ c^* e^{(- i \omega t)} v_{22} \right]  e^{(-\sigma t)}.\label{eq:z2SOL}
\end{eqnarray}
These equations show that $z_1$ and $z_2$ have the same decay rate ($\sigma$) but different oscillatory components with frequency $\omega$. These results are consistent with the POP analysis of \citet{SheshadriPlumb2017} who showed that EOF1 and EOF2 are, respectively, the real and imaginary parts of a single decaying-oscillatory POP mode (see their Section~4b). As a results, the two modes have the same decay rate and frequency, but have different auto-correlation function decay rates and have strong lag cross-correlations because the oscillations are out of phase. A key contribution of our work is to find the decay rate $\sigma$ and frequency $\omega$ as a function of $b_{jk}$ and $\tau_j$ (Eqs.~(\ref{eq:sigma})-(\ref{eq:omega})).

\begin{figure*}[t]
\centering
\includegraphics[width=29pc,angle=0,trim={2cm 15cm 1cm 4cm},clip]{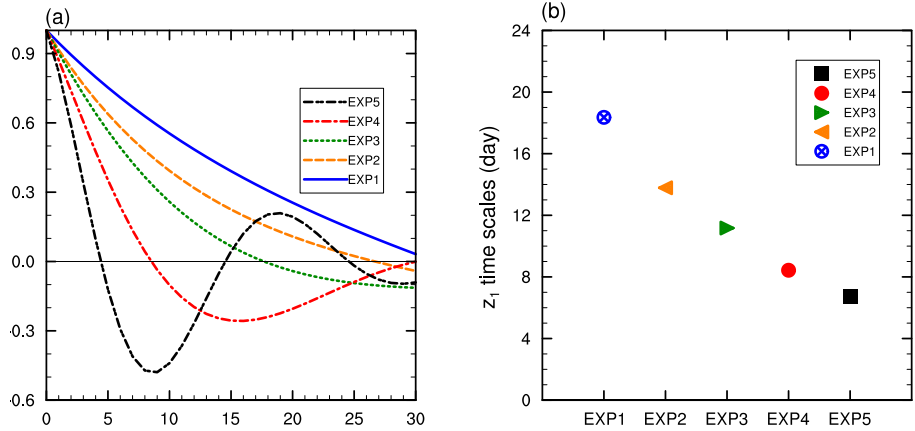}
\caption{Auto-correlation functions of ${z_1}$ (a) and their corresponding $e$-folding decorrelation time scales (b) from the analytical solutions for the experiment with no cross-EOF feedback (EXP1) and the experiments with increasing cross-EOF feedback strength (EXP2-EXP5). The prescribed feedback strength $b_{jk}$ are shown in Table~\ref{tab:3}.}
\label{fig:11}
\end{figure*}

To understand the effects of the feedback strength $b_{jk}$ on the persistence of $z_j$, we compute the analytical solutions for 5 systems that have the same $b_{11}>0$ and $b_{22}=0$ (Table~\ref{tab:3}): In EXP1, there is no cross-EOF feedback ($b_{12}=b_{21}=0$), while in EXP2-EXP5, $b_{12}>0$ and $b_{21}<0$ and they have been doubled from experiment to experiment. Figure~\ref{fig:11} shows the auto-correlation coefficients of $z_{1}$ and their $e$-folding decorrelation time scales for EXP1-EXP5. EXP1, corresponding to non-propagating regimes, has the slowest-decaying auto-correlation function, i.e., longest $e$-folding decorrelation time scale (Figs.~\ref{fig:11}a,b). EXP2-EXP5, which all satisfy condition (Eq.~\ref{eq:condition1}), have faster-decaying auto-correlation functions, i.e., shorter $e$-folding decorrelation time scale, consistent with our earlier results in idealized GCM and stochastic prototype (Figs. 4 and 10). As discussed above, in the propagating regime, the eigenvectors and the corresponding eigenvalue are complex and thus, $z_{1,2}$ do not decay just exponentially, but rather show some oscillatory characteristics too (Fig.~\ref{fig:11}a, Eqs.~(\ref{eq:z1SOL})-(\ref{eq:z2SOL})). Since the frequency of these oscillations $\omega$ (Eq.~(\ref{eq:omega})) increases as the cross-EOF feedback strengths increase, shorter time scales in $z_{1}$ are expected in the experiment with stronger $b_{12}b_{21}$ (Fig.~\ref{fig:11}b).

\begin{table}
\caption{Prescribed feedback strengths (in day$^{-1}$) used to analyze the impact of cross-EOF feedbacks on the decorrelation time scales of $z_1$ and $z_2$. The imposed damping rates of friction are $\tau_1$=$\tau_2$= $8$~days.}
\begin{center}
\begin{tabular}{ccccccrrcrcrcrc}
\topline
Feedback               & $b_{11} $          & $b_{12}$  	& $b_{21}$ 	                & $b_{22}$ \\
\midline
Exp1                       & 0.040              & 0.000 		& 0.000			        & 0.000\\
Exp2	                      & 0.040              & 0.060 		& -0.025			        & 0.000\\
Exp3                       & 0.040              & 0.120 		& -0.050			        & 0.000\\
Exp4                       & 0.040              & 0.240		 & -0.100				& 0.000\\
Exp5                       & 0.040              & 0.480	        & -0.200				& 0.000\\
\botline
\end{tabular} \label{tab:3}
\end{center}
\end{table}

\begin{figure*}[t]
\centering
\includegraphics[width=19pc,angle=90,trim={1cm 1cm 1cm 1cm},clip]{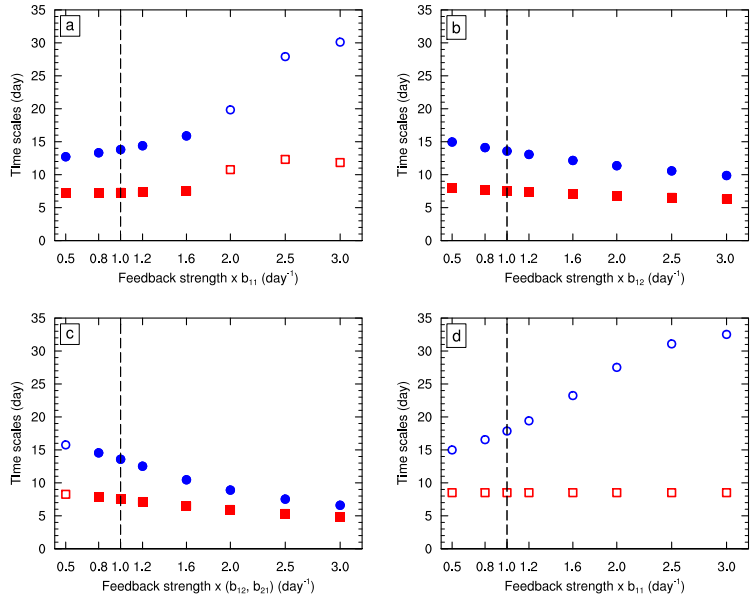}
\caption{The computed $e$-folding decorrelation time scale (day) of $z_1$ (blue circles) and $z_2$ (red squares) as a function of feedback strengths (day$^{-1}$). The impact of varying (a) $b_{11}$, (b) $b_{12}$, and (c) $b_{12}$ and $b_{21}$ on the decorrelation time scale (the $y$-axis) while all other $b_{jk}$ are kept the same. The $x$-axis shows the value of varied $b_{jk}$ as fraction of the value in EXP2 (Table~\ref{tab:3}); the vertical dashed line indicates the control values. (d) The impact of varying $b_{11}$ in EXP1 (Table~\ref{tab:3}). The filled indicates that the parameters satisfy the condition for propagating regimes, i.e., existence of decaying-oscillatory solutions (Eq.~(\ref{eq:condition1})).}
\label{fig:12}
\end{figure*}

The dependence of the $e$-folding decorrelation time scales of $z_1$ and $z_2$ on the feedback strengths, and in particular the cross-EOF feedback strengths, is further evaluated in Fig.~\ref{fig:12}. In Fig.~\ref{fig:12}a, it is clearly seen that the impact of increasing $b_{11}>0$ in the propagating regime (filled symbols) is to increase the persistence, i.e., decorrelation time scale, of $z_1$ (Fig.~\ref{fig:12}a), consistent with  increasing the positive eddy-zonal flow feedback ($z_1$-onto-$z_1$ through $m_1$). However, when the feedback is further increased to twice the control value, condition (\ref{eq:condition1}) for the existence of a decaying-oscillatory solution is not satisfied anymore, and consistent with this, we see that the system undergoes a transition to the non-propagating regime. Further increasing $b_{11}$ leads to substantially more persistent $z_1$ and less persistence $z_2$. Note that in non-propagating regimes when $b_{12}b_{21} \neq 0$, the decay of $z_2$ depends on $b_{11}$ too (see Eq.~(\ref{eq:lambda})).

Figure~\ref{fig:12}b shows that in the propagating regime, unlike increasing $b_{11}>0$, increasing $b_{12}>0$ leads to reduction in the persistence of $z_1$. This is the counter-intuitive behavior we had observed earlier in the stochastic prototype (Section~3\ref{sec:32}). Now we understand that this is because increasing $b_{12}$ increases the frequency of the oscillation $\omega$ in the system, resulting in reduction in the the decorrelation time scale of $z_1$ (and $z_2$); also see Fig.~\ref{fig:11}. Such impact can even be more pronounced when both cross-EOF feedbacks $b_{12}$ and $b_{21}$ are increased (Fig.~\ref{fig:12}c), leading to shorter decorrelation time scales. Because a positive $b_{12}$ decreases the persistence of $z_1$, we do not refer to is as a "positive feedback". To understand this behavior, we have to keep in mind that in the eddy forcing of $z_1$ ($z_2$), i.e., $m_1$ in Eq.~(\ref{eq:m1}) ($m_2$ in Eq.~(\ref{eq:m2})), $b_{12}>0$ ($b_{21}<0$) is the coefficient of $z_2$ ($z_1$). When $z_2$ leads $z_1$, they are negatively correlated (Figs.~\ref{fig:4}b, \ref{fig:7}b, and \ref{fig:10}b), thus $z_2$ multiplied by $b_{12}>0$ reduces $m_1$ that is forcing $z_1$, decreasing the persistence of $z_1$. Similarly, when $z_1$ leads $z_2$, they are positively correlated, thus $z_1$ multiplied by $b_{21}<0$ reduces $m_2$ and thus the persistence of $z_2$.

Finally, for the sake of completeness, we also examine the effect of increasing $b_{11}$ in the absence of cross-EOF feedback (Fig.~\ref{fig:12}d). As expected increasing  $b_{11}$ leads to increasing the persistence of $z_1$ and has no impact on the persistence of $z_2$ as now $z_1$ and $z_2$ are completely decoupled.

\subsection{Quantifying eddy-zonal flow feedbacks in reanalysis and idealized GCM}
\label{sec:34}

The results of Sections~3\ref{sec:32} and ~3\ref{sec:33} show the importance of carefully quantifying and interpreting the eddy-zonal flow feedbacks, including the cross-EOF feedbacks, to understand the variability of the zonal-mean flow. 

Table~\ref{tab:4} presents the feedback strengths obtained from applying (\ref{eq:b11})-(\ref{eq:b22}) with $l=8-20$ days to the year-round SH reanalysis data. We find $b_{11} = 0.038 $~day$^{-1}$, a positive feedback from $z_1$ onto $z_1$, consistent with the findings of \citet{Lorenz2001} in their pioneering work. This estimate of $b_{11}$ is slightly higher than what we find using the single-EOF approach ($b_{11} = 0.035 $~day$^{-1}$), which is the same as what \citet{Lorenz2001} found using their spectral cross-correlation method. We also find non-zero cross-EOF feedbacks: $b_{12}=0.059$~day$^{-1}$ and $b_{21}=-0.020$~day$^{-1}$. We also estimate $b_{22}=0.017$~day$^{-1}$ that is slightly higher from what the single-EOF approach yields (Table~\ref{tab:4}). The estimated feedback strengths and friction rates ($\tau$) in Table~\ref{tab:4} satisfy the condition for propagating regime (Eq.~\ref{eq:condition1}). It should be noted that we also extended our approach to include the leading 3 EOFs and quantified the 9 feedback strengths; however, we found the effects of EOF3 on EOF1 and EOF2 negligible, which suggests that a two-EOF model (\ref{eq:b11})-(\ref{eq:b22}) is enough to describe the current SH large-scale circulation (not shown).      

\begin{table}
\caption{Feedback strengths (in day$^{-1}$) estimated for year-round ERA-Interim reanalysis. The damping rates of friction are estimated as $\tau_1=8.3$~days and $\tau_2=8.4$~days following the methodology in Appendix~A of \citet{Lorenz2001}.}
\begin{center}
\begin{tabular}{ccccccrrcrcrcrc}
\topline
Feedback                  		    	 & $b_{11} $   & $b_{12}$  	  & $b_{21}$ 	  & $b_{22}$ \\
\midline
Eqs.~(\ref{eq:b11})-(\ref{eq:b22})       & 0.038        & 0.059 		   & -0.020		  & 0.017    \\
LH01                               		         & 0.035	    & - 			   & -			  & 0.002    \\
\botline
\end{tabular} \label{tab:4}
\end{center}
\end{table}

Table~\ref{tab:5} presents the feedback strengths obtained from applying (\ref{eq:b11})-(\ref{eq:b22}) with $l=8-20$ days to the two setups of the idealized GCM. In the non-propagating regime, we find $b_{11} = 0.133$~day$^{-1}$, and small $b_{22}$ and negligible $b_{12}$ and $b_{21}$, indicating the absence of cross-EOF feedbacks, consistent with insignificant $m_1z_2$ and $m_2z_1$ cross-correlations (Figs.~\ref{fig:2}e-f). The values of $b_{jk}$ do not satisfy the condition for propagating regime, which is consistent with weak cross-correlation between $z_1$ and $z_2$ at long lags (Fig.~\ref{fig:2}b). These results suggest that a strong $z_1$-onto-$z_1$ feedback dominates the dynamics of the annular mode in this setup (the standard Held-Suarez configuration), which leads to an unrealistically persistent annular mode, similar to what is seen in Fig.~\ref{fig:12}d, and consistent with the findings of previous studies \citep{SonLee2006,SonLee2008,MaHassanzadehKuang2017}. Using the linear response function (LRF) of this setup \citep{HassanzadehKuang2016,HassanzadehKuang2019} showed that this eddy-zonal flow feedback is due to enhanced low-level baroclinicity (as proposed by \citet{Robinson2000} and \citet{Lorenz2001}) and estimated, from a budget analysis, that the positive feedback is increasing the persistence of the annular mode by a factor of two.

In the propagating regime, we find $b_{11} = 0.101$~day$^{-1}$, which is slightly lower than $b_{11}$ of the non-propagating regime. However, in the propagating regime, we also find strong cross-EOF feedbacks $b_{12}=0.075$~day$^{-1}$, $b_{21}=-0.043$~day$^{-1}$ as well as $b_{22}=0.023$~day$^{-1}$. These feedback strengths satisfy the condition for propagating regime, consistent with strong cross-correlation between $z_1$ and $z_2$ at long lags (Fig.~\ref{fig:4}b). Comparing the two rows of Table~\ref{tab:5} and Figs.~\ref{fig:2}a and \ref{fig:4}a with Table~\ref{tab:4} and Fig.~\ref{fig:7}a suggests that while it is true that the $b_{11}$ of the the idealized GCM's non-propagating regime is larger than that of the SH reanalysis (by a factor of 3.5), the unrealistic persistence of $z_1$ in this setup (time scale $\approx 65$~days) compared to that of the reanalysis (time scale $\approx 10$~days; compare Figs.~\ref{fig:2}a and \ref{fig:7}a) could be, at least partially, due to the absence of cross-EOF feedbacks (thus oscillations), which as we showed earlier in Section~3c, reduce the persistence of the annular modes. The GCM setup with propagating regime has $b_{11}$ that is around 2.7 times larger than that of the SH reanalysis, yet their $z_1$ $e$-folding decorrelation time scales are comparable (14~days vs. 10~days).

\begin{table}
\caption{Feedback strengths (in day$^{-1}$) estimated for the idealized GCM setups with non-propagating and propagating regimes. The estimated damping rates of friction are $\tau_1$=7.4 days and $\tau_2$=7.6 days for the GCM setup with non-propagating regime, and $\tau_1$=7.1 days and $\tau_2$=7.4 days for the GCM setup with propagating regime (estimated using the methodology in Appendix~A of \citet{Lorenz2001}).}
\begin{center}
\begin{tabular}{ccccccrrcrcrcrc}
\topline
Feedback                  		   & $b_{11} $       & $b_{12}$  	   & $b_{21}$ 	        & $b_{22}$   \\
\midline
Non-propagating                           & 0.133		    & 0.003        	   & 0.002                  & 0.021	 \\
Propagating                 	            & 0.101             & 0.075		   & -0.043			& 0.023      \\
\botline
\end{tabular} \label{tab:5}
\end{center}
\end{table}

These findings show the importance of quantifying and examining cross-EOF feedbacks to fully understand the dynamics and variability of the annular modes and to better evaluate how well the GCMs simulate the extratopical large-scale circulation.

\section{Concluding remarks}
\label{sec:4}

The low-frequency variability of the extra-tropical large-scale circulation is often studied using a reduced-order model of the leading EOF of zonal-mean zonal wind. The key component of this model (LH01) is an internal-to-troposphere eddy-zonal flow interaction mechanism which leads to a positive feedback of EOF1 onto itself, thus increasing the persistence of the annular mode \citep{Lorenz2001}. However, several studies have showed that under some circumstances, strong couplings exist between EOF1 and EOF2 at some lag times, resulting in decaying-oscillatory, or propagating, annular modes (e.g. \citealt{SonLee2006,SonLee2008,SheshadriPlumb2017}). In the current study, following the methodology of \citet{Lorenz2001} and using data from the SH reanalysis and two setups of an idealized GCM that produce circulations with a dominant non-propagating or propagating regime, we first show strong cross-correlations between EOF1 (EOF2) and the eddy forcing of EOF2 (EOF1) at long lags, suggesting that cross-EOF feedbacks might exist in the propagating regimes. These findings together demonstrate that there is a need to extend the single-EOF model of LH01 and build a model that includes, at a minimum, both leading EOFs and accounts for their cross feedbacks. 

With similar assumptions and simplifications used in \citet{Lorenz2001}, we have developed a two-EOF model for propagating annular modes (consisting of a system of two coupled ODEs, Eqs.~(\ref{eq:z1})-(\ref{eq:z2}) with (\ref{eq:m1})-(\ref{eq:m2})) that can account for the cross-EOF feedbacks. In this model, the strength of the feedback of $k$th EOF onto $j$th EOF is $b_{jk}$ ($j,k=1,2$). Using the analytical solution of this model, we derive conditions for the existence of the propagating regime based on the feedback strengths. It is shown that the propagating regime, which requires a decaying-oscillatory solution of the coupled ODEs, can exist only if the cross-EOF feedbacks have opposite signs ($b_{12}  b_{21} <0$), and if and only if the following criterion is satisfied: $\left ( b_{11}-b_{22} \right )^2  <  -4 b_{12}  b_{21}$. These criteria show that non-zero cross-EOF feedbacks are essential components of the propagating regime dynamics.

Using this model and the idealized GCM and a stochastic prototype, we further show that cross-EOF feedbacks play an important role in controlling the persistence of the propagating annular modes (i.e., the $e$-folding decorrelation time scale of the zonal index, $z_j$) by setting the frequency of the oscillation $\omega $ (Eq.~(\ref{eq:omega})). Therefore, in this regime, the persistence of the annular mode (EOF1) does not only depend on the feedback of EOF1 onto itself, but also depends on the cross-EOF feedbacks. We find that as a result of the oscillation, the stronger the cross-EOF feedbacks, the less persistent the annular mode.

Applying the coupled-EOF model to the reanalysis data shows the existence of strong cross-EOF feedbacks in the current SH extratropical large-scale circulation. Annular modes have been found to be too persistent compared to observations in GCMs including IPCC AR4 and CMIP5 models \citep{Gerber2007,GerberPolvani2008,Bracegirdle2020}. This long persistence has been often attributed to a too strong positive EOF1-onto-EOF1 feedback in the GCMs. The dynamics and strength of this feedback depends on factors such as the mean flow and surface friction \citep{Robinson2000,Lorenz2001,ChenPlumb2009,HassanzadehKuang2019}. External (to troposphere) influence, e.g., from the stratospheric polar vortex, has been also suggested to affect the persistence of the annular modes \citep{Byrne2016,Saggioro2019}. Our results show that the cross-EOF feedbacks play an important role in the dynamics of the annular modes, and in particular, that their absence or weak amplitudes can increase the persistence, offering another explanation for the too-persistent annular modes in GCMs.

Overall, our findings demonstrate that to fully understand the dynamics of the large-scale extratropical circulation and the reason(s) behind the too-persistent annular modes in GCMs, the coupling of the leading EOFs and the cross-EOF feedbacks should be examine using models such as the one introduced in this study.

An important next step is to investigate the underlying dynamics of the cross-EOF feedbacks. So far we have pointed out that cross-EOF feedbacks are essential components of the propagating annular modes; however, the propagation itself is likely essential for the existence of cross-EOF feedbacks. In fact, our preliminary result shows that the cross-EOF feedbacks result from the out-of-phase oscillations of EOF1 (north-south jet displacement) and EOF2 (jet pulsation) leading to an orchestrated combination of equatorward propagation of wave activity (a baroclinic process) and nonlinear wave breaking (a barotropic process), which altogether act to reduce the total eddy forcings (not shown). In ongoing work, we aim to explain and quantify the propagating annular modes dynamics using the LRF framework of \citet{hassanzadeh2016linear2,HassanzadehKuang2016} and finite-amplitude wave-activity framework \citep{NakamuraZhu2010,LubisHuang2018,LubisHuangNakamura2018} that have been proven useful in understanding the dynamics of the non-propagating annular modes \citep{Nie2014,MaHassanzadehKuang2017,HassanzadehKuang2019}.

\acknowledgments
We thank Aditi Sheshadri, Ding Ma, and Orli Lachmy for insightful discussions. This work is supported by National Science Foundation (NSF) Grant AGS-1921413. Computational resources were provided by XSEDE (allocation ATM170020), NCAR's CISL (allocation URIC0004), and Rice University Center for Research Computing.



\appendix[A]
\appendixtitle{Standard Errors of Cross-Correlations using Bartlett's Formula}

Assuming two stationary normal time series $\left \{X_t  \right \}$ and $\left \{Y_t  \right \}$ ($t \in [0 \;\; T]$) with the corresponding auto-correlation functions $\rho_X(l)$ and $\rho_Y(l)$ and zero true cross-correlations, the standard error of the estimated cross-correlation at lag $l$ ($r_{XY} (l)$) can be computed as (see \citealt{Bartlett1978}, page 352):  
\begin{equation} 
\textup{var}\left \{r_{XY} (l) \right \}  =\frac{1}{T -  | l | } \sum_{g=-\infty }^{\infty} \left [ \rho_X(g) \rho_Y(g) \right ].
\end{equation} 
The null hypothesis  is $r_{XY} (l) = 0$, and it is rejected at the 5\% significance level if the estimated cross-correlation value at lag $l$ is larger than two times square root of the estimated standard error, i.e., $\left |r_{XY} (l) \right | > 2 \times \sqrt{\textup{var}\left \{r_{XY} (l) \right \}}$.


\bibliographystyle{ametsoc2014}
\bibliography{references}

\end{document}